\documentclass[cmp]{svjour}
\usepackage{amssymb,amsmath}

\spnewtheorem*{thm}{Theorem}{\bf}{\it}
\spnewtheorem*{pf}{Proof}{\bf}{\rm}
\numberwithin{equation}{section}

\newcommand{\wt}{\tilde}
\newcommand{\wh}{\hat}
\newcommand{\wb}{\bar}
\renewcommand{\th}[1]{\wh{\wt{#1}}}
\newcommand{\hb}[1]{\wb{\wh{#1}}}
\newcommand{\bt}[1]{\wt{\wb{#1}}}
\newcommand{\thb}[1]{\wh{\wt{\wb{#1}}}}

\newcommand{\cQ}{\mathcal{Q}}
\newcommand{\cR}{\mathcal{R}}
\newcommand{\cs}{\mathcal{S}}

\newcommand{\oT}{\mathsf{T}}
\newcommand{\oA}{\mathsf{A}}
\newcommand{\oB}{\mathsf{B}}
\newcommand{\oX}{\mathsf{X}}

\newcommand{\ssp}{\mathfrak{p}}
\newcommand{\ssq}{\mathfrak{q}}
\newcommand{\sst}{\mathfrak{t}}
\newcommand{\ssl}{\mathfrak{l}}
\newcommand{\sse}{\mathfrak{e}}
\newcommand{\ssf}{\mathfrak{f}}

\newcommand{\sn}{\mathrm{sn}}
\newcommand{\cn}{\mathrm{cn}}
\newcommand{\dn}{\mathrm{dn}}
\newcommand{\ETA}{\mathrm{H}}
\newcommand{\THE}{\mathrm{\Theta}}
\newcommand{\GAM}{\mathrm{\Gamma}}
\newcommand{\LAM}{\mathrm{\Lambda}}

\newcommand{\dd}{{*}}
\newcommand{\oh}{{\frac{1}{2}}}
\newcommand{\sIm}[1]{\stackrel{#1}{\sim}}
\newcommand{\ti}{\kappa}
\newcommand{\iu}{i}

\title{A constructive approach to the soliton solutions of integrable quadrilateral lattice equations}
\titlerunning{A constructive approach to the soliton solutions of lattice equations}
\author{James Atkinson\inst{1}\and Frank Nijhoff\inst{2}}
\institute{School of Mathematics and Statistics, The University of Sydney, NSW 2006, Australia.\\ \email{jamesa@maths.usyd.edu.au}
\and School of Mathematics, The University of Leeds, Leeds, LS2 9JT, UK.\\ \email{frank@maths.leeds.ac.uk}}
\authorrunning{J Atkinson and F W Nijhoff}
\date{September 25, 2009}

\begin{document}
\maketitle
\begin{abstract}
Scalar multidimensionally consistent quadrilateral lattice equations are studied.
We explore a confluence between the superposition principle for solutions related by the B\"acklund transformation, and the method of solving a Riccati map by exploiting two known particular solutions.
This leads to an expression for the $N$-soliton-type solutions of a generic equation within this class. 
As a particular instance we give an explicit $N$-soliton solution for the primary model, which is Adler's lattice equation (or $Q4$).
\end{abstract}
\section{Introduction}\label{INTRO}
Hirota's $N$-soliton solution of the Korteweg-de Vries (KdV) equation \cite{hir1,hir2} was characterised in terms of a B\"acklund transformation originally by Wahlquist and Estabrook in \cite{we}.
Nowadays the close relationship between soliton solutions and B\"acklund transformations is a part of the landscape of soliton theory.

The discrete systems which form the natural analogues of integrable partial differential equations of KdV-type are the scalar multidimensionally consistent quadrilateral lattice equations.
Amongst the features which are more transparent on the level of the discrete equations is the B\"acklund transformation, in fact a natural auto-B\"acklund transformation is inherent in the discrete equation itself because of the multidimensional consistency.
This is helpful in defining the very notion of soliton solution in the discrete setting, where the usual physical properties attributed to these solutions are less well studied.
It also makes the solutions arising from iterative application of the B\"acklund transformation very well suited to a treatment on the discrete level.
Particularly (we have found) when combined with an elementary technique for solving the Riccati equation, which we now take a moment to explain.

In the scalar case the discrete Riccati equation ({\it Riccati map}) is an equation of the form
\begin{equation}
\cR(u,\wt{u}) := a_0 + a_1 u + a_2 \wt{u} + a_3 u \wt{u} = 0,\label{gr}
\end{equation}
where $u=u(n)$ and $\wt{u}=u(n+1)$ are values of the dependent variable $u$ as a function of the independent variable $n\in\mathbb{Z}$, and in general the equation is non-autonomous so that the {\it coefficients} $a_0\ldots a_3\in\mathbb{C}$ are functions of $n$.
One of the many interesting properties of this equation is as follows: given two particular solutions $v=v(n)$ and $w=w(n)$, the substitution
\begin{equation}
u = \frac{v-\eta w}{1-\eta}\label{gs}
\end{equation}
reduces (\ref{gr}) to a homogeneous linear equation for the new variable $\eta=\eta(n)$.
Specifically 
\begin{equation}
\wt{\eta} = -\frac{\cR(w,\wt{v})}{\cR(v,\wt{w})}\,\eta \quad \Rightarrow \quad \cR(u,\wt{u})=0\label{requiv}
\end{equation}
as a consequence of the vanishing of the first and last terms in the expansion
\begin{equation}
(1-\eta)(1-\wt{\eta})\cR(u,\wt{u}) = \cR(v,\wt{v}) - \eta \cR(w,\wt{v}) - \wt{\eta} \cR(v,\wt{w})+\eta\wt{\eta} \cR(w,\wt{w}).\label{rexpand}
\end{equation}
This technique has been applied to the Riccati map associated with the lattice B\"acklund transformation previously \cite{ahn1,nah}.
In the present communication we show that a similar technique applied to the superposition principle makes it possible to construct a closed-form expression for the solution found by the application of $N$ B\"acklund transformations.
In other words, it leads to an expression for the $N$-soliton-type solutions.

From classification results due to Adler, Bobenko and Suris (ABS) \cite{abs1,abs2} there has emerged a core list of canonical forms for the scalar multidimensionally consistent quadrilateral lattice equations.
With one notable exception, explicit $N$-soliton solutions have been found for all of the equations on the ABS list \cite{ahn2,nah,hz}, and their relation to the B\"acklund transformation has also been examined \cite{nah}.
The exception is the lattice equation discovered by Adler \cite{adl} (cf. also \cite{nij,as}), which acquired the name $Q4$ in \cite{abs1}.
$Q4$ holds the position as the primary model on the ABS list because all of the others can be found from this single equation by a process of degeneration.
The principal application of the technique developed here is therefore to $Q4$, for which an explicit $N$-soliton solution has not (to our knowledge) hitherto been given.

We start in Section \ref{CLASS} by specifying the class of equations under consideration.
In Section \ref{SIMPLIFIED} the B\"acklund transformation and superposition principle of a generic equation within this class are written in terms of the associated variable $\eta$.
The iterative application of the B\"acklund transformation in this new variable then leads to the main $N$-soliton formula in Section \ref{ITERATION}.
In Section \ref{A1SOL} we provide a simple illustrative example, using equation $A1$ from \cite{abs1}.
In Section \ref{Q4SOL} we use the $N$-soliton formula to exhibit an explicit $N$-soliton solution of $Q4$, this builds on the seed and one-soliton solution constructed for this equation previously in \cite{ahn1}.

\section{The class of equations}\label{CLASS}
We will study an equation of a generic form which encompasses all of those listed by ABS in \cite{abs1}.
Specifically, we consider a polynomial of degree one in four scalar variables,
\begin{equation}
\cQ_{p,q}(u,\wt{u},\wh{u},\th{u}) := a_0(p,q) + a_1(p,q)u + a_2(p,q)\wt{u} + \ldots + a_{15}(p,q)u\wt{u}\wh{u}\th{u},\label{al}
\end{equation}
in which the coefficients are functions of {\it lattice parameters} $p$ and $q$.
We assume the {\it Kleinian symmetry} and {\it covariance} of the polynomial,
\begin{gather}
\cQ_{p,q}(u,\wt{u},\wh{u},\th{u}) =  \cQ_{p,q}(\wt{u},u,\th{u},\wh{u}) = \cQ_{p,q}(\wh{u},\th{u},u,\wt{u})\label{klein},\\
\cQ_{p,q}(u,\wt{u},\wh{u},\th{u}) = -\cQ_{q,p}(u,\wh{u},\wt{u},\th{u}),\label{covar}
\end{gather}
as well as compatibility of the system
\begin{equation}
\begin{array}{cc}
\cQ_{p,q}(u,\wt{u},\wh{u},\th{u})=0,\quad& \cQ_{p,q}(\wb{u},\bt{u},\hb{u},\thb{u})=0,\\
\cQ_{q,r}(u,\wh{u},\wb{u},\hb{u})=0,\quad& \cQ_{q,r}(\wt{u},\th{u},\bt{u},\thb{u})=0,\\
\cQ_{r,p}(u,\wb{u},\wt{u},\bt{u})=0,\quad& \cQ_{r,p}(\wh{u},\hb{u},\th{u},\thb{u})=0,
\end{array}
\label{concon}
\end{equation}
for any $p,q$ and $r$.
By the compatibility of (\ref{concon}) we mean that given $u$, $\wt{u}$, $\wh{u}$, $\wb{u}$, $\th{u}$, $\hb{u}$ and $\bt{u}$ satisfying the equations on the left, the remaining equations on the right are all satisfied by the {\it same} value of $\thb{u}$.
It is from the polynomial $\cQ$ that we then define an autonomous {\it quadrilateral lattice equation},
\begin{equation}
\cQ_{p,q}(u,\wt{u},\wh{u},\th{u}) = 0,\label{ge}
\end{equation}
where $u=u(n,m)$, $\wt{u}=u(n+1,m)$, $\wh{u}=u(n,m+1)$ and $\th{u}=u(n+1,m+1)$ are now values of the dependent variable $u$ as a function of independent variables $n,m\in\mathbb{Z}$.
The covariance (\ref{covar}) and the compatibility of (\ref{concon}) make the lattice equation {\it multidimensionally consistent} \cite{nw,bs1}.\footnote{It should be expected that all polynomials satisfying (\ref{al}) and (\ref{klein}) admit parameterisation leading to the multidimensional consistency \cite{abs2,via}. On the other hand, for polynomials with no non-linear terms the assumption (\ref{klein}) seems less natural \cite{atk1}.}

\section{B\"acklund transformation and superposition principle}\label{SIMPLIFIED}
We are interested in the solutions of (\ref{ge}) constructed by the application of $N$ consecutive B\"acklund transformations to an initial {\it seed} solution.
To apply the Riccati solution technique we will presuppose that the seed solution admits {\it covariant extension}.
\begin{definition}\label{ce}
A covariantly extended solution of (\ref{ge}) is one which also satisfies the (compatible) equations
\begin{subequations}\label{as}
\begin{gather}
\cQ_{p,l_i}(u,\wt{u},\oT_i u,\oT_i \wt{u}) = 0,\quad
\cQ_{q,l_i}(u,\wh{u},\oT_i u,\oT_i \wh{u}) = 0,\label{as1}\\
\cQ_{l_i,l_j}(u,\oT_iu,\oT_ju,\oT_i\oT_ju) = 0,\label{as2}
\end{gather}
\end{subequations}
for all $i,j\in\{1\ldots N\}$.
\end{definition}
In this definition we have introduced shift operators $\oT_1\ldots \oT_N$ associated with new lattice parameters $l_1\ldots l_N$, so the requirement is that $u$ inhabits an $(N+2)$-dimensional lattice satisfying the equations (\ref{ge}) and (\ref{as}) throughout.
We remark that situations in which there is no {\it natural} covariant extension are conceivable.
However, covariant-extendibility is not a restriction because generically, given a solution of (\ref{ge}), the multidimensional consistency implies that solutions of the equations (\ref{as}) exist.\footnote{All solutions of (\ref{ge}) arising as $N$-cycles of the B\"acklund transformation - which are a discrete analogue of finite-gap solutions \cite{an,wei1,wei2}, {\it do} have a natural covariant extension (cf. the example in Section \ref{Q4SOL}).
On the other hand, for solutions which result from a generic initial value problem on the lattice, a natural covariant extension is not manifest.}

We give a definition of the (extended) B\"acklund transformation of (\ref{ge}).
\begin{definition}\label{bt}
When two functions $u$ and $v$ satisfy the system of equations
\begin{subequations}\label{BT}
\begin{gather}
\cQ_{p,r}(u,\wt{u},v,\wt{v}) = 0,\quad
\cQ_{q,r}(u,\wh{u},v,\wh{v}) = 0,\label{BTpq}\\
\cQ_{l_i,r}(u,\oT_iu,v,\oT_iv) = 0, \quad i\in\{1\ldots N\},\label{BTce}
\end{gather}
\end{subequations}
we say they are related by the B\"acklund transformation of (\ref{ge}) with B\"acklund parameter $r$, and write $u\sIm{r} v$.
\end{definition}
One usually considers only the two-dimensional system (\ref{BTpq}) in the definition of the B\"acklund transformation, but here we have included auxiliary equations so that the B\"acklund transformation preserves the covariant extension.
The following statement is locally equivalent to the assumption of multidimensional consistency described in Section \ref{CLASS}.
\begin{lemma}\label{bl1}
If $u$ is a covariantly extended solution of (\ref{ge}) then considered as a system of equations for $v$ the B\"acklund equations $u\sIm{r}v$ are compatible.
Moreover the function $v$ which emerges is again a covariantly extended solution of (\ref{ge}).
\end{lemma}
Now, the special choice of B\"acklund parameter $r\in\{l_1\ldots l_N\}$ causes an interplay between the B\"acklund transformation and the covariant extension, and it is this which facilitates the Riccati solution technique.
\begin{lemma}\label{fl}
Let $u$ be a covariantly extended solution of (\ref{ge}). 
Then for each $i\in\{1\ldots N\}$ the substitution
\begin{equation} u_i = \frac{[\oT_i^{-1} - \eta_i \oT_i]u}{1-\eta_i} \label{subi}\end{equation}
reduces the B\"acklund equations $u\sIm{l_i}u_i$ to the following compatible homogeneous linear system for the new variable $\eta_i$,
\begin{subequations}\label{etasys}
\begin{gather}
\wt{\eta}_i = -\frac{\cQ_{p,l_i}(u,\wt{u},\oT_iu,\oT^{-1}_i\wt{u})}{\cQ_{p,l_i}(u,\wt{u},\oT^{-1}_iu,\oT_i\wt{u})}\,\eta_i,\quad
\wh{\eta}_i = -\frac{\cQ_{q,l_i}(u,\wh{u},\oT_iu,\oT^{-1}_i\wh{u})}{\cQ_{q,l_i}(u,\wh{u},\oT^{-1}_iu,\oT_i\wh{u})}\,\eta_i,\label{etapq}\\
\oT_j\eta_i = -\frac{\cQ_{l_j,l_i}(u,\oT_j u,\oT_iu,\oT_j\oT^{-1}_iu)}{\cQ_{l_j,l_i}(u,\oT_ju,\oT^{-1}_iu,\oT_j\oT_iu)}\,\eta_i,\quad j\in\{1\ldots N\}.\label{etace}
\end{gather}
\end{subequations}
\end{lemma}
\begin{pf}
The covariant extension provides two particular solutions of the system $u\sIm{l_i} u_i$, namely $u_i=\oT_i u$ immediately from (\ref{as}), and $u_i=\oT^{-1}_i u$, which can be seen by applying $\oT_i^{-1}$ to (\ref{as}) and using the symmetry (\ref{klein}).
Now observe that, considered as equations for $v$, the B\"acklund transformation (\ref{BT}) is a system of compatible Riccati maps because of the assumption (\ref{al}). 
The substitution (\ref{subi}) can then be seen to bring $u\sIm{l_i}u_i$ to (\ref{etasys}) by considering an expansion of the form (\ref{rexpand}) on each equation.
The compatibility of (\ref{etasys}) is inherited.\qed
\end{pf}
We remark that the behaviour of $\eta_i$ and $u_i$ away from the origin in the lattice direction associated with shift $\oT_i$ is likely to be quite special. 
Actually in practice we will only use the covariant extension of $u_i$ and $\eta_i$ into the remaining lattice directions associated with shifts $\oT_j$ with $j\neq i$.

Let us turn now to consider the superposition principle for solutions of (\ref{ge}) related by the B\"acklund transformation. 
Again we state the following without proof because it is a standard calculation based on the multidimensional consistency.
\begin{lemma}\label{bl2}
Let $u$ be a covariantly extended solution of (\ref{ge}) and suppose $u_1$ and $u_2$ are such that $u\sIm{r_1}u_1$ and $u\sIm{r_2}u_2$ for some $r_1$ and $r_2$.
Then the function $u_{12}$ determined algebraically by the equation $\cQ_{r_1,r_2}(u,u_1,u_2,u_{12})=0$ satisfies the B\"acklund relations $u_1\sIm{r_2}u_{12}$ and $u_2\sIm{r_1}u_{12}$.
\end{lemma}
This brings us to the crux of the method, it lies in considering Lemma \ref{bl2} in the case that $r_1,r_2\in\{l_1\ldots l_N\}$.
The resulting interplay with the covariant extension (in which $l_1\ldots l_N$ play the role of lattice parameters, cf. Definition \ref{ce}) leads to a relation, derived from the superposition formula, connecting now $\eta$-variables associated with each B\"acklund transformation.
\begin{lemma}\label{ml}
Let $u$ be a covariantly extended solution of (\ref{ge}) and suppose the functions $\eta_1\ldots \eta_N$ are such that
\begin{equation}
u_i =\frac{[\oT_i^{-1}-\eta_i\oT_i]u}{1-\eta_i} \quad \Rightarrow \quad u\sIm{l_i}u_i, \qquad i\in\{1\ldots N\}.\label{imp1}
\end{equation}
Then for each $i,j\in\{1\ldots N\}$ the following identity holds
\begin{equation}
(\oT_j^{-1}\eta_i)(\oT_i\eta_j) = (\oT_i^{-1}\eta_j)(\oT_j\eta_i).\label{eta_property}
\end{equation}
Furthermore the new function $\eta_{ij}$ defined as
\begin{equation}
\eta_{ij} = \frac{\ 1-\oT_j\,\eta_i\ }{1\!-\!\oT_j^{-1}\eta_i}\,\oT_i^{-1}\eta_j\label{sp2}
\end{equation}
is such that
\begin{equation}
u_{ij} =\frac{[\oT_j^{-1}-\eta_{ij}\oT_j]u_i}{1-\eta_{ij}} \quad \Rightarrow \quad u_i\sIm{l_j}u_{ij}.\label{imp2}
\end{equation}
\end{lemma}
\begin{pf}
The proof is based on Lemma \ref{bl2} plus a calculation which exploits the assumptions made about the equation (\ref{ge}) described in Section \ref{CLASS}.
First we observe that 
\begin{equation}
\cQ_{l_i,l_j}(u,\oT_i^{\mu} u, \oT_j^{\nu} u, \oT_i^{\mu}\oT_j^{\nu} u) = 0, \quad \mu,\nu\in\{+1,-1\},\label{qij}
\end{equation}
for all $i,j\in\{1\ldots N\}$.
In the case $\mu=\nu=1$ this is immediate from the covariant extension (\ref{as2}), the other cases follow from the symmetry (\ref{klein}).
By combining (\ref{qij}) with the affine-linearity (\ref{al}) we have for each $\mu,\nu\in\{+1,-1\}$ and each pair $i,j\in\{1\ldots N\}$
\begin{equation}
\cQ_{l_i,l_j}(u,\oT_i^{\mu} u, \oT_j^{\nu} u, z)=\cs_{ij}(u,\oT_i^{\mu}u,\oT_j^{\nu}u)(z-\oT_i^\mu\oT_j^\nu u)
\label{qs}
\end{equation}
as an identity in $z$.
Here we have introduced a polynomial $\cs$ of degree one in three variables defined in terms of $\cQ$ as
\begin{equation}
\cs_{ij}(w,x,y) := \cQ_{l_i,l_j}(w,x,y,1)-\cQ_{l_i,l_j}(w,x,y,0) = \partial_z \cQ_{l_i,l_j}(w,x,y,z).
\end{equation}
This new polynomial inherits the symmetry
\begin{equation}
\cs_{ij}(w,x,y) = - \cs_{ji}(w,y,x)\label{ssym}
\end{equation}
from the covariance of $\cQ$ (\ref{covar}).
The simplification which occurs by using (\ref{qs}) to replace occurrences of $\cQ$ with $\cs$ will be the principal mechanism exploited in the calculation.

Now, according to Lemma \ref{fl} the assumption (\ref{imp1}) means that each $\eta_i$ satisfies (\ref{etasys}). 
Using (\ref{qs}) in (\ref{etace}) as well as in the equation found by applying $\oT_j^{-1}$ to (\ref{etace}), we are able to deduce the following simplified equations for shifts on $\eta_i$ into the extended lattice directions,
\begin{equation}
\oT_j\eta_i = \frac{\cs_{ij}(u,\oT_iu,\oT_ju)}{\cs_{ij}(u,\oT_i^{-1}u,\oT_ju)}\,\eta_i, \quad \oT^{-1}_j\eta_i = \frac{\cs_{ij}(u,\oT_iu,\oT^{-1}_ju)}{\cs_{ij}(u,\oT_i^{-1}u,\oT^{-1}_ju)}\,\eta_i,\label{etasimp}
\end{equation}
which hold for all pairs $i,j\in\{1\ldots N\}$.
The relations (\ref{etasimp}) combined with (\ref{ssym}) may be used to directly verify the property (\ref{eta_property}).

We now proceed with the principal calculation; substituting for $u_i$ and $u_j$ leads to the following expansion of the superposition formula
\begin{equation}
\begin{split}
&(1-\eta_i)(1-\eta_j)\cQ_{l_i,l_j}(u,u_i,u_j,u_{ij})\phantom{\Big{\vert}}\\
&= \cQ_{l_i,l_j}(u,\oT_i^{-1}u,\oT_j^{-1}u,u_{ij})-\eta_i\cQ_{l_i,l_j}(u,\oT_iu,\oT_j^{-1}u,u_{ij})\\
&\quad \ -\eta_j\cQ_{l_i,l_j}(u,\oT_i^{-1}u,\oT_ju,u_{ij})+\eta_i\eta_j\cQ_{l_i,l_j}(u,\oT_iu,\oT_ju,u_{ij}),\\
&= \cs_{ij}(u,\oT_i^{-1}u,\oT_j^{-1}u)(u_{ij}-\oT_i^{-1}\oT_j^{-1}u)-\eta_i\cs_{ij}(u,\oT_iu,\oT_j^{-1}u)(u_{ij}-\oT_i\oT_j^{-1}u)\\
&\quad \ -\eta_j\cs_{ij}(u,\oT_i^{-1}u,\oT_ju)(u_{ij}-\oT_i^{-1}\oT_ju)+\eta_i\eta_j\cs_{ij}(u,\oT_iu,\oT_ju)(u_{ij}-\oT_i\oT_ju),\\
&= \cs_{ij}(u,\oT_i^{-1}u,\oT_j^{-1}u)\big((u_{ij}-\oT_i^{-1}\oT_j^{-1}u)-(\oT_j^{-1}\eta_i)(u_{ij}-\oT_i\oT_j^{-1}u)\\
&\quad \ -(\oT_i^{-1}\eta_j)(u_{ij}-\oT_i^{-1}\oT_ju)+(\oT_j\eta_i)(\oT_i^{-1}\eta_j)(u_{ij}-\oT_i\oT_ju)\big).
\end{split}\label{ijexp}
\end{equation}
The first equality here results from substituting for $u_i$ and $u_j$ using the equation on the left of (\ref{imp1}) and exploiting the assumed affine-linearity of $\cQ$ (\ref{al}).
To get the second equality from the first we have used (\ref{qs}).
The third equality follows from the second by using (\ref{etasimp}) whilst bearing in mind (\ref{ssym}).

The expansion (\ref{ijexp}) thus enables us to write an expression for the (covariantly extended) solution of (\ref{ge}) determined from $u$, $u_i$ and $u_j$ by the superposition formula $\cQ_{l_i,l_j}(u,u_i,u_j,u_{ij})=0$, 
\begin{equation}
u_{ij} = \frac{\big[\oT_i^{-1}\oT_j^{-1}-(\oT_j^{-1}\eta_i)\oT_i\oT_j^{-1}-(\oT_i^{-1}\eta_j)\oT_i^{-1}\oT_j+(\oT_j\eta_i)(\oT_i^{-1}\eta_j)\oT_i\oT_j\big]u}{1-(\oT_j^{-1}\eta_i)-(\oT_i^{-1}\eta_j)+(\oT_j\eta_i)(\oT_i^{-1}\eta_j)}.\label{duij}
\end{equation}
It is easily verified that the expression for $\eta_{ij}$ in (\ref{sp2}) leads to equality between $u_{ij}$ given on the left of (\ref{imp2}) and $u_{ij}$ found by superposition in (\ref{duij}).
The statement (\ref{imp2}) therefore follows from Lemma \ref{bl2}.\qed
\end{pf}
Thus it is demonstrated that, beyond what might be expected, the function $\eta_{ij}$ is determined completely in terms of $\eta_i$ and $\eta_j$ by an expression (\ref{sp2}) which does not depend on the particular form of the polynomial $\cQ$.
The additional statement of Lemma \ref{ml}, (\ref{eta_property}), can be viewed as encoding the permutability of the B\"acklund transformations (specifically $u_{ij}=u_{ji}$ in (\ref{imp2})) as a property of the $\eta$-variables; this property will also play a key role later.

\section{B\"acklund iteration}\label{ITERATION}
Here we apply $N$ consecutive B\"acklund transformations to a covariantly extended seed solution of (\ref{ge}).
Choosing the B\"acklund parameters to coincide with the lattice parameters $l_1\ldots l_N$ of the covariant extension (cf. Definition \ref{ce}) renders the B\"acklund iteration scheme tractable and yields a closed-form expression for the resulting $N$-soliton-type solution.
In the main theorem which follows it is both convenient and (as will be seen) natural to introduce a solution $v$ and functions $\phi_1\ldots \phi_N$ which are shifted versions of the seed solution $u$ and the variables $\eta_1\ldots \eta_N$ of Section \ref{SIMPLIFIED}.
\begin{thm}
Given a covariantly extended solution of (\ref{ge}) denoted by $u$, let 
\begin{equation}
v = \big[{\textstyle \prod_{i=1}^N\oT_i^{-1}}\big]u \label{vdef}
\end{equation}
and for each $i\in\{1\ldots N\}$ suppose the function $\phi_i$ satisfies the system
\begin{subequations}\label{phisys}
\begin{gather}
\wt{\phi}_i = -\frac{\cQ_{p,l_i}(\oT_iv,\oT_i\wt{v},\oT^2_iv,\wt{v})}{\cQ_{p,l_i}(\oT_iv,\oT_i\wt{v},v,\oT_i^2\wt{v})}\,\phi_i,\quad
\wh{\phi}_i = -\frac{\cQ_{q,l_i}(\oT_iv,\oT_i\wh{v},\oT^2_iv,\wh{v})}{\cQ_{q,l_i}(\oT_iv,\oT_i\wh{v},v,\oT_i^2\wh{v})}\,\phi_i,\label{phipq}\\
\oT_j\phi_i = -\frac{\cQ_{l_j,l_i}(\oT_iv,\oT_i\oT_j v,\oT_i^2v,\oT_jv)}{\cQ_{l_j,l_i}(\oT_iv,\oT_i\oT_jv,v,\oT_i^2\oT_jv)}\,\phi_i, \quad j\in\{1\ldots N\}.\label{phice}
\end{gather}
\end{subequations}
Then the linear difference operators $\oB_1\ldots\oB_N$ defined as
\begin{equation}
\oB_i:=1-\phi_i\oT_i^2, \quad i\in\{1\ldots N\},\label{Bdef}
\end{equation}
commute amongst each other, and the function $u^{(N)}$ defined in terms of these operators by the expression
\begin{equation}
u^{(N)}:=\frac{\big[{\textstyle \prod_{i=1}^N\oB_i}\big]v}{\big[{\textstyle \prod_{i=1}^N\oB_i}\big]1},
\label{nss}
\end{equation}
is a solution of (\ref{ge}) related to $u$ by the composition of $N$ B\"acklund transformations with B\"acklund parameters $l_1\ldots l_N$.
\end{thm}
\begin{pf}
Let us begin by supposing $\eta_i$ satisfies (\ref{etasys}) for each $i\in\{1\ldots N\}$.
Then according to Lemma \ref{fl} the function $u_i$ defined by (\ref{subi}) satisfies $u\sIm{l_i}u_i$.
In other words the following is true when we take $i=0$: 
\begin{equation}
u_{1\ldots ij} = \frac{[\oT_{j}^{-1}-\eta_{1\ldots ij}\oT_j]u_{1\ldots i}}{1-\eta_{1\ldots ij}} \quad \Rightarrow \quad u_{1\ldots i}\sIm{l_j}u_{1\ldots ij}, \qquad j\in\{1\ldots N\}.\label{uit}
\end{equation}
Now, using Lemma \ref{ml}, if we define the functions $\eta_{1\ldots ij}$ recursively by the expression
\begin{equation}
\eta_{1\ldots ij}=\frac{1-\oT_j\eta_{1\ldots i}}{1-\oT_j^{-1}\eta_{1\ldots i}}\,\oT_i^{-1}\eta_{1\ldots (i-1)j}, \quad i,j\in\{1\ldots N\},\label{etait}
\end{equation}
then (\ref{uit}) is true for any $i=\kappa\in\{1\ldots N\}$ as a consequence of it being true when $i=\kappa-1$.

The recurrence (\ref{etait}) is not difficult to solve.
Let us exhibit the first few iterations, writing them in the following way (which requires use of (\ref{eta_property})) makes the emerging general formula quite evident:
\begin{equation*}
\begin{split}
\eta_{1j} \! & = \! \frac{\oT_j[\oT_1^{-1}\!\!-\!\eta_1\oT_1]1}{\oT_j^{-1}[\oT_1^{-1}\!\!-\!\eta_1\oT_1]1}\oT_1^{-1}\eta_j,\\
\eta_{12j} \! & = \! \frac{\oT_j[\oT_2^{-1}\!\!-\!(\oT_1^{-1}\eta_2)\oT_2][\oT_1^{-1}\!\!-\!\eta_1\oT_1]1}{\oT_j^{-1}[\oT_2^{-1}\!\!-\!(\oT_1^{-1}\eta_2)\oT_2][\oT_1^{-1}\!\!-\!\eta_1\oT_1]1}\oT_1^{-1}\oT_2^{-1}\eta_j,\\
\eta_{123j} \! & = \! \frac{\oT_j[\oT_3^{-1}\!\!-\!(\oT_1^{-1}\oT_2^{-1}\eta_3)\oT_3][\oT_2^{-1}\!\!-\!(\oT_1^{-1}\eta_2)\oT_2][\oT_1^{-1}\!\!-\!\eta_1\oT_1]1}{\oT_j^{-1}[\oT_3^{-1}\!\!-\!(\oT_1^{-1}\oT_2^{-1}\eta_3)\oT_3][\oT_2^{-1}\!\!-\!(\oT_1^{-1}\eta_2)\oT_2][\oT_1^{-1}\!\!-\!\eta_1\oT_1]1\!}\oT_1^{-1}\oT_2^{-1}\oT_3^{-1}\eta_j.
\end{split}\label{etaseq}
\end{equation*}
To write the $i^{th}$ term in this sequence we introduce linear difference operators
\begin{equation}
\oA_i := \prod_{j}^{i\curvearrowleft 1}\left[\oT_j^{-1}-\big(\big[\textstyle{\prod_{k=1}^{j-1}\oT_k^{-1}}\big]\eta_j\big)\oT_j\right], \quad i\in\{1\ldots N\},\label{Adef}
\end{equation}
in terms of which
\begin{equation}
\eta_{1\ldots ij} = \frac{\oT_j\oA_i\,1}{\oT_j^{-1}\oA_i\,1}\,\big[{\textstyle \prod_{k=1}^{i}\oT_k^{-1}}\big]\eta_j.\label{etasol}
\end{equation}
This can be verified as the solution of (\ref{etait}) directly by substitution.

We now reconstruct the solutions using the equation on the left of (\ref{uit}), in fact it is sufficient to assume $i$ and $j$ are consecutive, so we need only consider the recurrence
\begin{equation}
u_{1\ldots i} = \frac{[\oT_i^{-1}-\eta_{1\ldots i}\oT_i]u_{1\ldots (i-1)}}{1-\eta_{1\ldots i}}, \quad i\in\{1\ldots N\}.
\label{uit2}
\end{equation}
The solution is simply
\begin{equation}
u_{1\ldots i} = \frac{\oA_iu}{\oA_i1}, \quad i\in\{1\ldots N\}\label{usol}
\end{equation}
which, bearing in mind (\ref{etasol}), can be easily verified by substitution.
Thus we have found an expression for solutions $u_{1\ldots i}$ which satisfy the desired B\"acklund relations $u_{1\ldots (i-1)}\sIm{l_i}u_{1\ldots i}$, $i\in\{1\ldots N\}$.

Now, we began the proof by assuming that, for each $i\in\{1\ldots N\}$, $\eta_i$ satisfied (\ref{etasys}).
Actually if we make the association
\begin{equation}
\eta_i = \big[{\textstyle \prod_{j=1,j\neq i}^N \oT_j}\big]\phi_i,\label{etaphi}
\end{equation}
then direct substitution shows that (\ref{etasys}) is just a consequence of (\ref{phisys}) and (\ref{vdef}) which occurred in the hypotheses of the theorem.
Using the association (\ref{etaphi}) it is straightforward to see that
\begin{equation}
\oA_N = \big[{\textstyle \prod_{i=1}^N\oB_i}\big]\big[{\textstyle \prod_{i=1}^N\oT_i^{-1}}\big].\label{ABrel}
\end{equation}
The commutativity of the operators $\oB_1\ldots \oB_N$, which enables us to remove the ordering of their product in (\ref{ABrel}), follows from the identity
\begin{equation}
\phi_i(\oT_i^2\phi_j) = \phi_j(\oT_j^2\phi_i),
\end{equation}
which is just (\ref{eta_property}) expressed in the shifted variables through (\ref{etaphi}).
From (\ref{ABrel}) and (\ref{vdef}) it is clear that $u^{(N)}$ in (\ref{nss}) is simply $u_{1\ldots N}$ in (\ref{usol}) expressed in the shifted variables, which completes the proof.\qed
\end{pf}

We conclude this section with a few remarks.
The functions $\phi_1\ldots \phi_N$ arise in the commuting operators defined in (\ref{Bdef}) making it natural to adopt these variables in preference to $\eta_1\ldots \eta_N$.
The function $v$ in (\ref{vdef}) is one of $2^N$ particular solutions of the iterative B\"acklund scheme which are provided by the covariant extension.
The others are manifest in (\ref{nss}) when the functions $\phi_1\ldots \phi_N$ are taken to be the trivial solutions of (\ref{phisys}), namely constant at 0 or $\infty$, implying $u^{(N)}\in\{[\oT_1^{-1}\cdots\oT_N^{-1}]u=v,[\oT_1\oT_2^{-1}\cdots \oT_N^{-1}]u,\ldots,[\oT_1\cdots \oT_N]u\}$.

Note that the function $u^{(N)}$ in (\ref{nss}) lies on an $(N+2)$-dimensional lattice, the desired $N$-soliton-type solution as a function of $n$ and $m$ alone is realised by discarding the covariant extension, i.e., by evaluation at the {\it origin} in the additional lattice directions associated with the shifts $\oT_1\ldots\oT_N$.

For the examples, as will be seen in the following two sections, it is useful to introduce the function
\begin{equation}
f:=\big[{\textstyle \prod_{i=1}^N\oB_i}\big]1,\label{fdef}
\end{equation}
which appears in the denominator of the $N$-soliton formula (\ref{nss}).
The operator present here has the expansion
\begin{multline}
\prod_{i=1}^{N}\oB_i = 1 - \sum_{i=1}^N \phi_i\oT_i^2 + \sum_{i=1}^N\sum_{j=i+1}^N\phi_i(\oT_i^2\phi_j)\oT_i^2\oT_j^2\\ - \sum_{i=1}^N\sum_{j=i+1}^N\sum_{k=j+1}^N\phi_i(\oT_i^2\phi_j)(\oT_i^2\oT_j^2\phi_k)\oT_i^2\oT_j^2\oT_k^2+\ldots,\label{Bexp}
\end{multline}
so the function $f$ is reminiscent of the one introduced by Hirota \cite{hir1,hir2}.
Additionally, we observe that the function $f$ can be expressed in a factorised form
\begin{equation}
f = (1-\phi_{1\ldots N})(1-\phi_{1\ldots N-1}) \cdots (1-\phi_{12})(1-\phi_1)
\end{equation}
through functions $\phi_{1\ldots i}$, which, for each $i\in\{1\ldots N\}$, are related to $\eta_{1\ldots i}$ present in the proof (\ref{etasol}) by the relation $\phi_{1\ldots i} = \big[\prod_{j=i+1}^N\oT_j^{-1}\big]\eta_{1\ldots i}$.
Or in terms of the shifted variables alone
\begin{equation}
\phi_{1\ldots i} = \frac{\big[\textstyle \oT_i^2\prod_{j=1}^{i-1}\oB_j\big]1}{\big[\textstyle \prod_{j=1}^{i-1}\oB_j\big]1}\,\phi_i = 1-\frac{\big[\textstyle \prod_{j=1}^{i}\oB_j\big]1}{\big[\textstyle \prod_{j=1}^{i-1}\oB_j\big]1}, \quad i\in\{1\ldots N\}.
\end{equation}

\section{An explicit $N$-soliton solution of $A1$}\label{A1SOL}
In this section we provide an illustrative example of the main theorem, in particular the $N$-soliton formula (\ref{nss}).
The solution we consider was (up to a gauge transformation) discovered previously by other methods in \cite{nah}.

Consider the expression
\begin{equation}
\cQ_{p,q}(u,\wt{u},\wh{u},\th{u}) := \frac{(u+\wh{u})(\wt{u}+\th{u})}{p^2-a^2} - \frac{(u+\wt{u})(\wh{u}+\th{u})}{q^2-a^2} + \frac{\delta^2a^4(p^2-q^2)}{(p^2\!-\!a^2)^2(q^2\!-\!a^2)^2}\label{A1}
\end{equation}
where $a\in\mathbb{C}\setminus\{0\}$ and $\delta\in\mathbb{C}$ are arbitrary constants. 
This is a polynomial of the form (\ref{al}); direct calculation will verify that it satisfies all of the properties listed in Section \ref{CLASS}.
The equation defined by (\ref{A1}) is equivalent to $A1$ in \cite{abs1}, the only difference being a point transformation of the lattice parameters $p\rightarrow a^2/(p^2-a^2)$, $q\rightarrow a^2/(q^2-a^2)$, which is useful for discussing solutions \cite{nah}. 

The equation defined by (\ref{A1}) has an elementary solution, $u$, of the form
\begin{equation}
u = A\psi+B/\psi,\label{A1seed}
\end{equation}
where
\begin{equation}
\psi = \psi(n,m) := \left(\frac{a+p}{a-p}\right)^n\left(\frac{a+q}{a-q}\right)^m \label{psidef}
\end{equation}
and $A,B\in\mathbb{C}$ are constants subject to the constraint
\begin{equation}
AB = \left({\delta}/{4}\right)^2.\label{ABcond}
\end{equation}
We will take $u$ in (\ref{A1seed}) as a seed solution to start the B\"acklund chain.
This solution admits a natural covariant extension; observe that
\begin{equation}
\wt{\psi} = \frac{a+p}{a-p}\,\psi, \quad \wh{\psi} = \frac{a+q}{a-q}\,\psi,\label{psipq}
\end{equation}
and consider complementing these with the compatible equations
\begin{equation}
\oT_i \psi = \frac{a+l_i}{a-l_i}\,\psi, \quad i\in\{1\ldots N\},\label{psice}
\end{equation}
in which we have introduced new parameters $l_1\ldots l_N \in\mathbb{C}\setminus\{-a,a\}$.
The equations (\ref{psice}) extend $\psi$ to the $(N+2)$-dimensional lattice (note that the function appearing in (\ref{psidef}) should now be interpreted as the evaluation of $\psi$ at the origin of the extra directions).
Of course the solution $u$ defined in terms of $\psi$ by (\ref{A1seed}) also inhabits the $(N+2)$-dimensional lattice, and it is straightforward to see it satisfies the additional system of equations (\ref{as}).
Thus, for the quadrilateral lattice equation defined by the polynomial (\ref{A1}), we have given a solution (\ref{A1seed}) and its natural covariant extension through (\ref{psice}).

We are interested in the result of $N$ applications of the B\"acklund transformation to the seed solution $u$.
To apply the $N$-soliton formula (\ref{nss}) we should construct from $u$ the related solution $v$ and the functions $\phi_1\ldots\phi_N$.
The solution $v$ is given in terms of $u$ by (\ref{vdef}).
With $u$ as in (\ref{A1seed}) we use (\ref{psice}) to find
\begin{equation}
v=\big[\textstyle\prod_{i=1}^N\oT_i^{-1}\big] u = A'\psi + B'/\psi \label{A1v}
\end{equation}
where the new constants $A'$ and $B'$ are related to $A$ and $B$ by the equations
\begin{equation}
A'=A\prod_{i=1}^N\frac{a-l_i}{a+l_i},\quad B'=B\prod_{i=1}^N\frac{a+l_i}{a-l_i},
\end{equation}
but satisfy the same constraint (\ref{ABcond}).
So $u$ and $v$ are the same up to a change in the value of some constants present in the solution.
The other ingredient present in the $N$-soliton formula (\ref{nss}) is the set of functions $\phi_1\ldots\phi_N$, these are defined in terms of $v$ by the system (\ref{phisys}) for each $i\in\{1\ldots N\}$.
Direct substitution of (\ref{A1}) and (\ref{A1v}) into (\ref{phisys}) followed by the use of $A'B'=(\delta/4)^2$, (\ref{psipq}) and (\ref{psice}) yields the following system for each $\phi_i$,
\begin{subequations}\label{A1phisys}
\begin{gather}
\wt{\phi}_i = \frac{p-l_i}{p+l_i}\,\phi_i, \quad \wh{\phi}_i = \frac{q-l_i}{q+l_i}\,\phi_i,\label{A1phipq}\\
\oT_j\phi_i = \frac{l_j-l_i}{l_j+l_i}\,\phi_i, \quad j\in\{1\ldots N\}.\label{A1phice}
\end{gather}
\end{subequations}
Integrating the equations (\ref{A1phipq}) we can write
\begin{equation}
\phi_i=\phi_i(n,m) = \phi_{i,0}\left(\frac{p-l_i}{p+l_i}\right)^n\left(\frac{q-l_i}{q+l_i}\right)^m, \quad i\in\{1\ldots N\},\label{A1phi}
\end{equation}
where $\phi_{1,0}\ldots \phi_{N,0}$ are independent of $n$ and $m$.

So we have constructed the solution $v$ and functions $\phi_1\ldots\phi_N$ present in the $N$-soliton formula (\ref{nss}).
We proceed by using (\ref{psice}) and (\ref{A1phice}) to write (\ref{nss}) in terms of {\it un-shifted} functions $\psi$ and $\phi_1\ldots \phi_N$; first consider the function $f$ defined in (\ref{fdef}) which appears in the denominator of (\ref{nss}), we find
\begin{multline}
f=f(\phi_1,\ldots,\phi_N)=1-\sum_{i=1}^N \phi_i + \sum_{i=1}^N\sum_{j=i+1}^N\phi_i\phi_jX_{ij}^2 \\- \sum_{i=1}^N\sum_{j=i+1}^N\sum_{k=j+1}^N\phi_i\phi_j\phi_kX_{ij}^2X_{ik}^2X_{jk}^2 + \ldots + (-1)^N\prod_{i=1}^N\Big(\phi_i\prod_{j=i+1}^NX_{ij}^2\Big),\label{A1fdef}
\end{multline}
where 
\begin{equation}
X_{ij} := \frac{l_j-l_i}{l_j+l_i} = -X_{ji}
\end{equation}
are constants such that $\oT_j\phi_i=X_{ij}\phi_i$.
A similar consideration of the numerator in (\ref{nss}) yields, after collecting terms in $\psi$ and $1/\psi$,
\begin{equation}
u^{(N)} = A'\psi \frac{f^-}{f} + (B'/\psi)\frac{f^+}{f}\label{A1nss},
\end{equation}
where we have introduced the functions $f^\pm$ which are defined in terms of $f$ in (\ref{A1fdef}) by the equations
\begin{equation}
f^\pm=f^\pm(\phi_1,\ldots,\phi_N) = f(x_1^{\pm 2}\phi_1,\ldots,x_N^{\pm 2}\phi_N), \quad x_i = \frac{a-l_i}{a+l_i}, \quad i\in\{1\ldots N\}.\label{A1fdef2}
\end{equation}
So, through the functions $f$, $f^+$ and $f^-$ given in (\ref{A1fdef}) and (\ref{A1fdef2}), $u^{(N)}$ given in (\ref{A1nss}) is a rational expression in the un-shifted functions $\psi$ and $\phi_1\ldots \phi_N$. 
Taking these functions as in (\ref{psidef}) and (\ref{A1phi}) respectively, (\ref{A1nss}) thus gives the solution as a function of $n$ and $m$ alone.
This is the $N$-soliton solution which results from $N$ applications of the B\"acklund transformation to the seed solution $u$ given in (\ref{A1seed}).
It is worth observing that for (\ref{A1nss}) to be a {\it true} $N$-soliton solution, i.e., one containing $N$ constants of integration, it is necessary that $l_i\neq 0$ for all $i\in\{1\ldots N\}$ and that $l_i\neq \pm l_j$ for all distinct $i,j\in\{1\ldots N\}$ (these conditions are in addition to the one already mentioned, $l_i\not\in\{-a,a\}$ for all $i\in\{1\ldots N\}$, which was required for the covariant extension).

Arranging the solution as we have done in (\ref{A1nss}) is suggestive of further structure behind this expression.
This further structure has a natural interpretation in the contexts of the Cauchy-matrix approach described in \cite{nah} (where this solution was first given) and the Casorati approach \cite{ahn2,hz}.
To finish this section we describe how this structure should be interpreted in relation to the approach introduced in Sections \ref{SIMPLIFIED} and \ref{ITERATION}.
The main idea is to extend the functions $\phi_1\ldots\phi_N$ into one further {\it special} lattice direction. 
This direction is associated with lattice-parameter $a$ and we denote shifts in this direction by $\oT_a$,
\begin{equation}
\oT_a\phi_i = \frac{a-l_i}{a+l_i}\,\phi_i, \quad i\in\{1\ldots N\},\label{Tadef}
\end{equation}
which inspecting (\ref{A1fdef2}) enables us to write $f^\pm = \oT_a^{\pm 2} f$.
(Note that to be consistent $\psi$ should take values in $\{0,\infty\}$ away from the origin in this special direction, but this won't play a role in the present construction.)
The form of the solution (\ref{A1nss}) then motivates the introduction of a new linear difference operator
\begin{equation}
\oX:=A'\psi\oT_a^{-2} + (B'/\psi)\oT_a^{2}\label{A1F}
\end{equation}
for which the following facts can be verified
\begin{equation}
\oX1=v, \quad [\oB_i,\oX]=0, \quad i\in\{1\ldots N\}.\label{A1F2}
\end{equation}
The first is clear from the definition (\ref{A1F}) whilst bearing in mind (\ref{A1v}), the second follows from (\ref{A1F}) and the definition of the operators $\oB_1\ldots \oB_N$ in (\ref{nss}) because
\begin{equation}
\psi(\oT_a^{-2}\phi_i) = \phi_i(\oT_i^2\psi), \quad (1/\psi)(\oT_a^{2}\phi_i) = \phi_i(1/\oT_i^2\psi), \quad i\in\{1\ldots N\},
\end{equation}
which are immediate from (\ref{Tadef}) and (\ref{psice}).
Now, using the properties (\ref{A1F2}) in (\ref{nss}) and the definition (\ref{fdef}) we find
\begin{equation}
u^{(N)} = \frac{\oX f}{f}.\label{A1nss2}
\end{equation}
Thus, for this example, have uncovered the presence of an operator $\oX$ defined in (\ref{A1F}) which, due to the properties (\ref{A1F2}), enables the $N$-soliton formula (\ref{nss}) to be written in the simplified form (\ref{A1nss2}).

\section{An explicit $N$-soliton solution of $Q4$}\label{Q4SOL}
We now use the $N$-soliton formula (\ref{nss}) to give an explicit $N$-soliton solution for the equation listed as $Q4$ in \cite{abs1}.
The construction of a seed solution for $Q4$ is itself an interesting problem which was solved previously in \cite{ahn1}, where the ensuing one-soliton solution was also given. 
The full $N$-soliton solution which results from this seed solution is given here for the first time.
\subsection{The equation $Q4$}
We consider $Q4$ in the Jacobi form \cite{hie}, which is related by a change of variables to the Weierstrass form given originally by Adler \cite{adl} (and expressed more concisely in \cite{nij}).
It may be defined in terms of the polynomial
\begin{equation}
\cQ_{\ssp,\ssq}(u,\wt{u},\wh{u},\wh{\wt{u}}):=p(u\wt{u}+\wh{u}\wh{\wt{u}})-q(u\wh{u}+\wt{u}\wh{\wt{u}})-\frac{pQ-qP}{1-p^2q^2}\left(u\wh{\wt{u}}+\wt{u}\wh{u}-pq(1+u\wt{u}\wh{u}\wh{\wt{u}})\right).\label{Q4}
\end{equation}
In Section \ref{CLASS} we did not specify the set from which the lattice parameters were taken, here $\ssp=(p,P)$ and $\ssq=(q,Q)$ are points on an elliptic curve, $\ssp,\ssq\in\GAM$, 
\begin{equation}
\GAM=\GAM(k):=\left\{(x,X) \,\vert\, X^2 = 1+x^4-(k+1/k)x^2\right\},\label{gam}
\end{equation}
where $k\in\mathbb{C}\setminus\{-1,0,1\}$, the Jacobi elliptic modulus, is a fixed constant.
The properties listed in Section \ref{CLASS} may all be verified directly for the polynomial (\ref{Q4}).

In our consideration of solutions for $Q4$ a central role will be played by the natural product that turns $\GAM$ into an abelian group.
We recall some facts and establish some notation regarding this here.
To start with let us write the rational representation of the group product:
\begin{equation}
\ssp \cdot \ssq = \left( \frac{pQ+qP}{1-p^2q^2}, \frac{Pp(q^4-1) - Qq(p^4-1)}{(1-p^2q^2)(qP-pQ)} \right),
\label{gp}
\end{equation}
and note the group identity is the point $\sse = (0,1)$ and the inverse of a point $\ssp$ is $\ssp^{-1} = (-p,P)$.
This group structure on $\GAM$ is naturally parameterised through the Jacobi elliptic functions by introducing the mapping
\begin{equation}
\ssf: \ z \mapsto \left(\sqrt{k}\,\sn(z;k),\cn(z;k)\dn(z;k)\right).\label{ssfdef}
\end{equation}
By convention (see for example Chapter 5 of \cite{akh}) the primitive periods of the function $z \mapsto \sn(z;k)$ are denoted $4K$ and $2\iu K'$ (here $\iu$ denotes the imaginary unit).
The mapping $\ssf$ defined in (\ref{ssfdef}) is a bijection from the fundamental parallelogram in $\mathbb{C}$ with vertices $0$, $4K$, $2\iu K'$ and $4K+2\iu K'$ to the curve $\GAM$, it brings the group product (\ref{gp}) down to addition on the torus, $\ssf(y)\cdot\ssf(z) = \ssf(y+z)$.

Because the group is abelian, the subset of $\GAM$ defined as $\LAM := \{\ssp\in\GAM\,\vert\,\ssp^2=\sse\}$ is also a subgroup.
Explicitly
\begin{subequations}\label{lam}
\begin{gather}
\LAM = \left\{ (0,1), (0,-1), \left(1/\epsilon,-1/\epsilon^2\right)_{\epsilon\rightarrow 0}, \left(-1/\epsilon,1/\epsilon^2\right)_{\epsilon\rightarrow 0}\right\}, \\ \ssf^{-1}(\LAM) = \left\{0,2K,\iu K',2K+\iu K'\right\}.
\end{gather}
\end{subequations}
In what follows a technical, but important role will be played by this subgroup.
One reason for this is that if $\ssp\in\LAM$, $\ssq\in\LAM$ or $\ssp\cdot\ssq^{-1}\in\LAM$, then the polynomial (\ref{Q4}) is reducible (in the projective sense, i.e., the leading order term of the polynomial is reducible when the unbounded points are considered).
In particular this means that the B\"acklund transformation of the equation defined by (\ref{Q4}), with B\"acklund parameter chosen from $\LAM$ (\ref{lam}), reduces to one of the point symmetries
\begin{equation}
\left\{u\rightarrow u, u\rightarrow -u, u\rightarrow 1/u, u\rightarrow -1/u\right\}.
\end{equation}

Although its role here is not essential, the three-leg form of $Q4$, which was established in \cite{abs1} and given for the Jacobi variables in \cite{bs2}, will be useful.
Introducing the uniformizing variables $\alpha$, $\beta$ and $\xi$ by writing $\ssp=\ssf(\alpha)$, $\ssq=\ssf(\beta)$ and $u=\sqrt{k}\,\sn(\xi)$, we have the following identity
\begin{equation}
\begin{split}
&\cQ_{\ssp,\ssq}(u,\wt{u},\wh{u},\wh{\wt{u}}) =\frac{ k^{3}\sqrt{k}\,\sn(\xi;k)\sn(\alpha;k)\sn(\beta;k)\sn(\alpha-\beta;k)}{1-\gamma}\times \\ 
\Big(&[\sn(\xi+\alpha;k)-\sn(\wt{\xi};k)][\sn(\xi-\beta;k)-\sn(\wh{\xi};k)][\sn(\xi-\alpha+\beta;k)-\sn(\wh{\wt{\xi}};k)]\gamma\\  - &[\sn(\xi-\alpha;k)-\sn(\wt{\xi};k)][\sn(\xi+\beta;k)-\sn(\wh{\xi};k)][\sn(\xi+\alpha-\beta;k)-\sn(\wh{\wt{\xi}};k)]\Big),
\end{split}\label{tlf}
\end{equation}
in which we have introduced $\gamma$ given by
\begin{equation}
\begin{split}
\gamma &= \frac{1-k^2\sn(\alpha;k)\sn(\beta;k)\sn(\xi;k)\sn(\xi+\alpha-\beta;k)}{1-k^2\sn(\alpha;k)\sn(\beta;k)\sn(\xi;k)\sn(\xi-\alpha+\beta;k)},\\
&= \frac{\THE(\xi+\alpha;k)\THE(\xi-\beta;k)\THE(\xi-\alpha+\beta;k)}{\THE(\xi-\alpha;k)\THE(\xi+\beta;k)\THE(\xi+\alpha-\beta;k)}.
\end{split}\label{gchoice}
\end{equation}
The second way of writing $\gamma$ in (\ref{gchoice}) which involves the Jacobi $\THE$ function (see appendix) leads to the three-leg form, the equality between the two expressions for $\gamma$ is an elliptic function identity ((\ref{mi}) in the appendix).
Assuming the first expression for $\gamma$ in (\ref{gchoice}) the identity (\ref{tlf}) may be verified using only the addition formula encoded in (\ref{gp}).
\subsection{Construction of the solution $v$}\label{Q4SEED}
The solution for $Q4$ given in \cite{ahn1} was found as a `fixed-point' \cite{wei1,wei2} or `1-cycle' \cite{an} of the B\"acklund transformation.
That is, a solution which is related to itself, $v\sIm{\sst} v$ for some freely chosen B\"acklund parameter $\sst=(t,T)\in\GAM\setminus\LAM$.
In Definition \ref{bt} we introduced the {\it extended} B\"acklund transformation. 
However for the moment we adhere to \cite{ahn1} and use the un-extended version, so that the defining equations for $v$ are
\begin{equation}
\cQ_{\ssp,\sst}(v,\wt{v},v,\wt{v}) = 0, \quad \cQ_{\ssp,\sst}(v,\wh{v},v,\wh{v}) = 0.\label{btfp}
\end{equation}
These are compatible symmetric biquadratic correspondences, their solution is in terms of shifts on an elliptic curve which, whilst being again of Jacobi type, has an elliptic modulus different than the modulus $k$ associated with the equation.
Thus, in order to write the solution of (\ref{btfp}) we need to introduce a second elliptic modulus, $k_\dd$, associated with the {\it solution}, $v$.
The new modulus is defined in terms of the free parameter $\sst$ by the relation
\begin{equation}
k_\dd+\frac{1}{k_\dd} = 2\frac{1-T}{t^2}.\label{ksdef}
\end{equation}
In addition to the new modulus we also define a new mapping
\begin{equation}
\ssf_\dd: \ z \mapsto \left(\sqrt{k_\dd}\,\sn(z;k_\dd),\cn(z;k_\dd)\dn(z;k_\dd)\right),
\end{equation}
which is a bijection from the fundamental parallelogram in $\mathbb{C}$ with vertices $0$, $4K_\dd$, $2\iu K'_\dd$, and $4K_\dd + 2\iu K'_\dd$, to the new elliptic curve $\GAM_\dd=\GAM(k_\dd)$.
Note that in (\ref{gp}) we gave the rational representation of the group product on $\GAM$ independently of the modulus $k$, so the product on $\GAM_\dd$ has the same rational representation, and of course $\ssf_\dd(y)\cdot\ssf_\dd(z)=\ssf_\dd(y+z)$.

Now, in order to write the solution of (\ref{btfp}) explicitly, we also need to introduce a relation between $\GAM$ and $\GAM_\dd$.
Specifically
\begin{equation}
\delta := \Big\{\left(\left(p,P\right),\left(p_\dd,P_\dd\right)\right)\in\GAM\times\GAM_\dd \ \Big\vert \ p_\dd^2=p\,\frac{pT-tP}{1-p^2t^2}, P_\dd = \frac{1}{t}\Big(p-\frac{pT-tP}{1-p^2t^2}\Big) \Big\}.\label{reldef}
\end{equation}
The relation $\delta$ does not establish a bijection between the two curves, in fact a generic point on one curve is related to {\it two} points on the other.\footnote{Consideration of the definition (\ref{reldef}) reveals that $(\ssp,\ssp_\dd)\in\delta \Leftrightarrow (\sst\cdot\ssp^{-1},\ssp_{\dd})\in\delta \Leftrightarrow (\ssp,\ssp_{\dd}^{-1})\in\delta$, therefore $\delta$ establishes a bijection between the sets $\{\{\ssp,\ssp^{-1}\cdot\sst\}\,\vert\,\ssp\in\GAM\}$ and $\{\{\ssp_\dd,\ssp_\dd^{-1}\}\,\vert\,\ssp_\dd\in\GAM_\dd\}$.
The group structure on $\GAM_\dd$ turns the latter set here into an abelian 2-group \cite{bv}, so $\delta$ endows the former set with this structure as well.}
In order to write down the solution of (\ref{btfp}) we introduce two points in $\GAM_\dd$,
\begin{equation}
\ssp_\dd=(p_\dd,P_\dd)=\ssf_\dd(\alpha_\dd), \quad \ssq_\dd=(q_\dd,Q_\dd)=\ssf_\dd(\beta_\dd)\label{pqs}
\end{equation}
which are defined in terms of the lattice parameters $\ssp$ and $\ssq$ through $\delta$,
\begin{equation}
(\ssp,\ssp_\dd),(\ssq,\ssq_\dd)\in\delta.\label{pqd}
\end{equation}
The solution $v$ is then given by
\begin{equation} 
v=\sqrt{k_{\dd}}\,\sn(\xi_\dd;k_\dd), \quad \xi_\dd=\xi^\dd_0 + n\alpha_\dd + m\beta_\dd,\label{seed}
\end{equation}
where $\xi^\dd_0$ is an arbitrary constant.

That the function $v$ in (\ref{seed}) satisfies (\ref{btfp}) was demonstrated in \cite{ahn1}.
Here our main concern is that this function satisfies $\cQ_{\ssp,\ssq}(v,\wt{v},\wh{v},\th{v})=0$, i.e., is a solution of the equation defined by (\ref{Q4}).
This can actually be verified quite easily by first checking the following identity:
\begin{equation}
\cQ_{\ssp,\ssq}(u,\wt{u},\wh{u},\th{u}) = \frac{1}{2}\Big(\frac{p}{p_\dd\!}\!-\!\frac{q}{q_\dd\!}\Big)\cQ_{\ssp_\dd,\ssq_\dd^{-1}}(u,\wt{u},\wh{u},\th{u})+\frac{1}{2}\Big(\frac{p}{p_\dd\!}\!+\!\frac{q}{q_\dd\!}\Big)\cQ_{\ssp_\dd,\ssq_\dd}(u,\wt{u},\wh{u},\th{u}), 
\label{split}
\end{equation}
which holds on any function $u=u(n,m)$, relying only on (\ref{pqd}).
It turns out that each term in (\ref{split}) vanishes on the function $v$ given in (\ref{seed}).
This can be seen by using (\ref{tlf}), where the substitution $(u,\ssp,\ssq)\rightarrow (v,\ssp_\dd,\ssq_\dd^{\pm 1})$ leads to the identifications $(k,\alpha,\beta,\xi)\rightarrow (k_\dd,\alpha_\dd,\pm\beta_\dd,\xi_\dd)$, whilst from (\ref{seed}) $\wt{\xi}_\dd = \xi_\dd + \alpha_\dd$ and so on.
(The function $v$ is the `non-germinating seed' \cite{ahn1} or `singular' \cite{abs2} solution of the equations $\cQ_{\ssp_\dd,\ssq_\dd^{\pm 1}}(u,\wt{u},\wh{u},\th{u})=0$.)

To finish this subsection we give the covariant extension of the solution $v$, which is necessary in order to apply the $N$-soliton formula (\ref{nss}).
This is very natural because $v$ was constructed as a $1$-cycle of the B\"acklund transformation, in fact a covariantly extended solution emerges if we construct the $1$-cycle of the {\it extended} B\"acklund transformation (\ref{BT}).
This amounts to complementing the system (\ref{btfp}) with the equations
\begin{equation}
\cQ_{\ssl_i,\sst}(v,\oT_iv,v,\oT_iv)=0, \quad i\in\{1\ldots N\},\label{btfp2}
\end{equation}
where we have introduced the B\"acklund parameters $\ssl_i=(l_i,L_i)\in\GAM$, $i\in\{1\ldots N\}$ and the shifts $\oT_1\ldots \oT_N$.
The function $v$ in (\ref{seed}) may be extended to solve the system (\ref{btfp2}) explicitly by introducing the points $\ssl_i^\dd=(l_i^\dd,L_i^\dd)=\ssf_\dd(\lambda_i^\dd)$, $i\in\{1\ldots N\}$ such that
\begin{equation}
(\ssl_i,\ssl_i^\dd) \in \delta, \quad i\in\{1\ldots N\},\label{llsd}
\end{equation}
and complementing (\ref{seed}) with the (evidently compatible) equations
\begin{equation}
\oT_i \xi_\dd = \xi_\dd + \lambda_i^\dd, \quad i\in\{1\ldots N\}.\label{cexi}
\end{equation}
\subsection{Integration of the equations for $\phi_1\ldots \phi_N$}\label{Q4PHI}
In the previous subsection we have recalled from \cite{ahn1} an explicit solution, $v$ in (\ref{seed}), for the equation $Q4$, and given its natural covariant extension through (\ref{cexi}).
To apply the main theorem in Section \ref{ITERATION} it remains to give the functions $\phi_1\ldots\phi_N$ which we construct from $v$ by integration of the equations (\ref{phisys}).
It turns out that with $\cQ$ as in (\ref{Q4}) and $v$ as in (\ref{seed}), the system (\ref{phisys}) for $\phi_i$ may be reduced by the substitution 
\begin{equation}
\phi_i = \frac{\THE(\xi_\dd+2\lambda_i^\dd;k_\dd)}{\THE(\xi_\dd;k_\dd)}\,\rho_i
\label{phirho}
\end{equation}
to the following {\it autonomous} system for the new variable $\rho_i$\footnote{The $\rho$ variable introduced here is different from the one defined in \cite{ahn1}.}:
\begin{subequations}\label{rhosys}
\begin{gather}
\wt{\rho}_i = \bigg(\frac{p_\dd l_i-p l_i^\dd}{p_\dd l_i+p l_i^\dd}\bigg)\frac{\THE(\lambda_i^\dd-\alpha_\dd;k_\dd)}{\THE(\lambda_i^\dd+\alpha_\dd;k_\dd)}\,\rho_i, \quad
\wh{\rho}_i = \bigg(\frac{q_\dd l_i-q l_i^\dd}{q_\dd l_i+q l_i^\dd}\bigg)\frac{\THE(\lambda_i^\dd-\beta_\dd;k_\dd)}{\THE(\lambda_i^\dd+\beta_\dd;k_\dd)}\,\rho_i, \label{rhopq} \\
\oT_j\rho_i = \bigg(\frac{l_j^\dd l_i-l_j l_i^\dd}{l_j^\dd l_i+l_j l_i^\dd}\bigg)\frac{\THE(\lambda_i^\dd-\lambda_j^\dd;k_\dd)}{\THE(\lambda_i^\dd+\lambda_j^\dd;k_\dd)}\,\rho_i, \quad j\in\{1\ldots N\}.\label{rhoce}
\end{gather}
\end{subequations}
We can integrate the equations (\ref{rhopq}), so that
\begin{equation}
\rho_i=\rho_{i,0}\left(\!\left(\frac{p_\dd l_i-p l_i^\dd}{p_\dd l_i+p l_i^\dd}\right)\!\frac{\THE(\lambda_i^\dd-\alpha_\dd;k_\dd)}{\THE(\lambda_i^\dd+\alpha_\dd;k_\dd)}\right)^n\!\left(\!\left(\frac{q_\dd l_i-q l_i^\dd}{q_\dd l_i+q l_i^\dd}\right)\!\frac{\THE(\lambda_i^\dd-\beta_\dd;k_\dd)}{\THE(\lambda_i^\dd+\beta_\dd;k_\dd)}\right)^m\label{rhonm}
\end{equation}
for $i\in\{1\ldots N\}$, where $\rho_{1,0}\ldots \rho_{N,0}$ are independent of $n$ and $m$.
Thus for each $i\in\{1\ldots N\}$ the function $\phi_i$ is given by (\ref{phirho}) in terms of $\xi_\dd$ (\ref{seed}), (\ref{cexi}) and the new function $\rho_i$ (\ref{rhonm}), (\ref{rhoce}).

It remains to satisfy the reader that the substitution (\ref{phirho}) reduces the system (\ref{phisys}) to the autonomous system (\ref{rhosys}).
Let us focus on the first equation of (\ref{phipq}), the other equations present in (\ref{phisys}) are similar.
We begin by using the identity (\ref{split}) which leads to the following reformulation of the first equation in (\ref{phipq}) 
\begin{equation}
\begin{split}
\wt{\phi}_i &= -\frac{\cQ_{\ssp,\ssl_i}(\oT_iv,\oT_i\wt{v},\oT_i^2 v, \wt{v})}{\cQ_{\ssp,\ssl_i}(\oT_iv,\oT_i\wt{v},v,\oT_i^2\wt{v})}\,\phi_i\\
&= \left(\frac{p_\dd l_i-pl_i^\dd}{p_\dd l_i+pl_i^\dd}\right)\frac{\cQ_{\ssp_\dd,{\ssl_i^\dd}^{-1}}(\oT_iv,\oT_i\wt{v},\oT_i^2 v, \wt{v})}{\cQ_{\ssp_\dd,\ssl_i^\dd}(\oT_iv,\oT_i\wt{v},v,\oT_i^2\wt{v})}\,\phi_i.
\end{split}\label{red1}
\end{equation}
That $\cQ_{\ssp_\dd,\ssl_i^\dd}(\oT_iv,\oT_i\wt{v},\oT_i^2v,\wt{v})=0$ and $\cQ_{\ssp_\dd,{\ssl_i^\dd}^{-1}}(\oT_iv,\oT_i\wt{v},v,\oT_i^2\wt{v})=0$ has also been exploited here, these are verifiable using the expansion (\ref{tlf}).
We then propose the following string of equalities
\begin{equation}
\begin{split}
&\frac{\cQ_{\ssp_\dd,{\ssl_i^\dd}^{-1}}(\oT_iv,\oT_i\wt{v},\oT_i^2 v, \wt{v})}{\cQ_{\ssp_\dd,\ssl_i^\dd}(\oT_iv,\oT_i\wt{v},v,\oT_i^2\wt{v})} \\
& \qquad = \frac{1-k_\dd^2\sn(\lambda_i^\dd;k_\dd)\sn(\lambda_i^\dd+\alpha_\dd;k_\dd)\sn(\xi_\dd+\lambda_i^\dd;k_\dd)\sn(\xi_\dd+\alpha_\dd+\lambda_i^\dd;k_\dd)}{1-k_\dd^2\sn(\lambda_i^\dd;k_\dd)\sn(\lambda_i^\dd-\alpha_\dd;k_\dd)\sn(\xi_\dd+\lambda_i^\dd;k_\dd)\sn(\xi_\dd+\alpha_\dd+\lambda_i^\dd;k_\dd)},\\
& \qquad = \frac{\THE(\lambda_i^\dd-\alpha_\dd;k_\dd)\THE(\xi_\dd+2\lambda_i^\dd+\alpha_\dd;k_\dd)\THE(\xi_\dd;k_\dd)}{\THE(\lambda_i^\dd+\alpha_\dd;k_\dd)\THE(\xi_\dd+\alpha_\dd;k_\dd)\THE(\xi_\dd+2\lambda_i^\dd;k_\dd)},\\
& \qquad = \frac{\THE(\lambda_i^\dd-\alpha_\dd;k_\dd)\THE(\wt{\xi}_\dd+2\lambda_i^\dd;k_\dd)\THE(\xi_\dd;k_\dd)}{\THE(\lambda_i^\dd+\alpha_\dd;k_\dd)\THE(\wt{\xi}_\dd;k_\dd)\THE(\xi_\dd+2\lambda_i^\dd;k_\dd)}.
\end{split}\label{red2}
\end{equation}
The first equality here can be verified using (\ref{tlf}) (modulo some manipulation).
The second follows from the first by an addition formula ((\ref{mi}) listed in the appendix).
The third equality is immediate from the second because $\wt{\xi}_\dd=\xi_\dd+\alpha_\dd$ by (\ref{seed}).
Finally, combining (\ref{red2}) with (\ref{red1}) motivates the substitution (\ref{phirho}) and clearly results in the first equation of (\ref{rhopq}).
\subsection{The $N$-soliton solution}\label{Q4SOLSOL}
In Sections \ref{Q4SEED} and \ref{Q4PHI} we have given the ingredients for the $N$-soliton formula (\ref{nss}), that is a solution $v$, its covariant extension, and the functions $\phi_1\ldots \phi_N$.
To write down the $N$-soliton solution constructed from these ingredients we consider first the function $f$ defined in (\ref{fdef}) which appears in the denominator of (\ref{nss}).
Substituting (\ref{phirho}) and (\ref{seed}) into (\ref{fdef}) whilst bearing in mind the expansion (\ref{Bexp}), and subsequently using (\ref{rhoce}) and (\ref{cexi}) to write the result in terms of the un-shifted functions $\xi_\dd$ and $\rho_1\ldots\rho_N$, we find
\begin{multline}
f = 1 - \sum_{i=1}^N \frac{\THE(\xi_\dd+2\lambda_i^\dd;k_\dd)}{\THE(\xi_\dd;k_\dd)}\rho_i + \sum_{i=1}^N\sum_{j=i+1}^N\frac{\THE(\xi_\dd+2\lambda_i^\dd+2\lambda_j^\dd;k_\dd)}{\THE(\xi_\dd;k_\dd)}\rho_i\rho_jX_{ij}^2\\ - \sum_{i=1}^N\sum_{j=i+1}^N\sum_{\ti=j+1}^N\frac{\THE(\xi_\dd+2\lambda_i^\dd+2\lambda_j^\dd+2\lambda_\ti^\dd;k_\dd)}{\THE(\xi_\dd;k_\dd)}\rho_i\rho_j\rho_\ti X_{ij}^2X_{i\ti}^2X_{j\ti}^2 + \ldots,\label{Q4f}
\end{multline}
where we have introduced the constants
\begin{equation}
X_{ij} := \bigg(\frac{l_j^\dd l_i-l_j l_i^\dd}{l_j^\dd l_i+l_j l_i^\dd}\bigg)\frac{\THE(\lambda_i^\dd-\lambda_j^\dd;k_\dd)}{\THE(\lambda_i^\dd+\lambda_j^\dd;k_\dd)} =
         \bigg(\frac{l_i^\dd l_j^\dd-l_i l_j}{l_i^\dd l_j^\dd+l_i l_j}\bigg)\frac{\ETA(\lambda_i^\dd-\lambda_j^\dd;k_\dd)}{\ETA(\lambda_i^\dd+\lambda_j^\dd;k_\dd)}
\label{Xdef}
\end{equation}
for each $i,j\in\{1\ldots N\}$. 
These constants appear because $\oT_j\rho_i=X_{ij}\rho_i$ as can be seen by comparing (\ref{Xdef}) and (\ref{rhoce}).
The equality between the alternative forms we have given for $X_{ij}$ in (\ref{Xdef}) can be verified using (\ref{set}) together with the addition formula encoded in (\ref{gp}) and the relations (\ref{llsd}).
A similar consideration of the numerator in the $N$-soliton formula (\ref{nss}), whilst bearing in mind (\ref{set}), makes it natural to introduce a new function $g$,
\begin{multline}
g = 1 - \sum_{i=1}^N \frac{\ETA(\xi_\dd+2\lambda_i^\dd;k_\dd)}{\ETA(\xi_\dd;k_\dd)}\rho_i + \sum_{i=1}^N\sum_{j=i+1}^N\frac{\ETA(\xi_\dd+2\lambda_i^\dd+2\lambda_j^\dd;k_\dd)}{\ETA(\xi_\dd;k_\dd)}\rho_i\rho_jX_{ij}^2 \\ - \sum_{i=1}^N\sum_{j=i+1}^N\sum_{\ti=j+1}^N\frac{\ETA(\xi_\dd+2\lambda_i^\dd+2\lambda_j^\dd+2\lambda_\ti^\dd;k_\dd)}{\ETA(\xi_\dd;k_\dd)}\rho_i\rho_j\rho_\ti X_{ij}^2X_{i\ti}^2X_{j\ti}^2 + \ldots .\label{Q4g}
\end{multline}
The $N$-soliton solution for the equation defined by (\ref{Q4}) may then be expressed as
\begin{equation}
u^{(N)}=v\frac{g}{f}\label{Q4nss}
\end{equation}
with $v$, $f$ and $g$ as in (\ref{seed}), (\ref{Q4f}) and (\ref{Q4g}), explicitly for $N\in\{0,1,2\}$ this is
\begin{align*}
&u^{(0)}=\frac{\ETA(\xi_\dd;k_\dd)}{\THE(\xi_\dd;k_\dd)},\quad
u^{(1)}=\frac{\ETA(\xi_\dd;k_\dd)\!-\!\ETA(\xi_\dd\!+\!2\lambda_1^\dd;k_\dd)\rho_1}{\THE(\xi_\dd;k_\dd)\!-\!\THE(\xi_\dd\!+\!2\lambda_1^\dd;k_\dd)\rho_1},\quad
u^{(2)}=\\&\frac{\ETA(\xi_\dd;k_\dd)\!-\!\ETA(\xi_\dd\!+\!2\lambda_1^\dd;k_\dd)\rho_1\!-\!\ETA(\xi_\dd\!+\!2\lambda_2^\dd;k_\dd)\rho_2\!+\!\ETA(\xi_\dd\!+\!2\lambda_1^\dd\!+\!2\lambda_2^\dd;k_\dd)\rho_1\rho_2 X_{12}^2}{\THE(\xi_\dd;k_\dd)\!-\!\THE(\xi_\dd\!+\!2\lambda_1^\dd;k_\dd)\rho_1\!-\!\THE(\xi_\dd\!+\!2\lambda_2^\dd;k_\dd)\rho_2\!+\!\THE(\xi_\dd\!+\!2\lambda_1^\dd\!+\!2\lambda_2^\dd;k_\dd)\rho_1\rho_2X_{12}^2}.
\end{align*}
The functions $\xi_\dd$ and $\rho_1\ldots\rho_N$ appearing in the solution are given in (\ref{seed}) and (\ref{rhonm}). 
The modulus $k$ and the constants $\ssp=(p,P)$ and $\ssq=(q,Q)$ taken from $\GAM=\GAM(k)$ defined in (\ref{gam}) are fixed (they appear in the equation), whereas the constants $\sst=(t,T)\in\GAM\setminus\LAM$ and $\ssl_i=(l_i,L_i)=\ssf(\lambda_i)\in\GAM$, $\rho_{i,0}\in\mathbb{C}\cup\{\infty\}$ for $i\in\{1\ldots N\}$ may all be chosen freely.
The modulus $k_\dd$ is determined from $\sst$ by (\ref{ksdef}), whilst the parameters $\ssp_\dd=(p_\dd,P_\dd)=\ssf_\dd(\alpha_\dd)$, $\ssq_\dd=(q_\dd,Q_\dd)=\ssf_\dd(\beta_\dd)$ and $\ssl_i^\dd=(l_i^\dd,L_i^\dd)=\ssf_\dd(\lambda_i^\dd)$ for $i\in\{1\ldots N\}$ which lie in $\GAM_\dd=\GAM(k_\dd)$ are all determined from their counterparts in $\GAM$ by the relation $\delta$ defined in (\ref{reldef}), $(\ssp,\ssp_\dd),(\ssq,\ssq_\dd),(\ssl_1,\ssl^\dd_1)\ldots(\ssl_N,\ssl^\dd_N)\in\delta$.

To ensure that the $N$-soliton solution (\ref{Q4nss}) has distinct functional dependence on each of the $N$ constants of integration $\rho_{1,0}\ldots\rho_{N,0}$ appearing in (\ref{rhonm}), and so constitutes a {\it true} $N$-soliton solution, we should take some care in how we choose the parameters $\ssl_1\ldots \ssl_N$.
First of all we should choose $\ssl_1\ldots\ssl_N\in\GAM\setminus(\LAM\cup\LAM\cdot\sst)$, this restriction ensures that, for each $i\in\{1\ldots N\}$, the B\"acklund equations associated with parameter $\ssl_i$ are not reducible, and also that the covariant extension provides two {\it distinct} solutions of the B\"acklund equations applied to the solution $v$, i.e., $\oT_iv\neq\oT^{-1}_iv$.
Furthermore, because we construct higher soliton solutions by superposition, we should also stipulate that $\ssl_i\cdot\ssl_j^{-1}\not\in\LAM$ for all distinct $i,j\in\{1\ldots N\}$.
Let us remark though, that soliton solutions for which $\ssl_i\in\LAM\cdot\sst$ for some $i\in\{1\ldots N\}$ or $\ssl_i\cdot\ssl_j^{-1}\in\LAM$ for some distinct $i,j\in\{1\ldots N\}$, are constructable by quadrature\footnote{Although the main results here rely on there being two, a {\it single} particular solution is in fact sufficient to subsequently solve a Riccati equation by quadrature.}, or recoverable from the solution given here by taking a subtle limit.
\subsection{Special lattice directions}
Rather than avoiding them altogether, we will now make particular use of the elements in $\LAM$.
It is quite straightforward to see that $\GAM\cap\GAM_\dd=\LAM$, so $\LAM$ is a subgroup of both $\GAM$ {\it and} $\GAM_\dd$\footnote{This is closely related to a characterisation of the deformation $\GAM\rightarrow\GAM_\dd$ which was established in \cite{atk2}.}.
It turns out that the relation $\delta$ (\ref{reldef}) is compatible with this shared subgroup structure on the two curves, $\ssp\in\LAM \Rightarrow (\ssp,\ssp)\in\delta$.
We will focus on the element $\sse\in\LAM$ as well as one other element which for convenience we denote by $\ssl_\oh$,
\begin{equation}
\ssl_\oh := \left(1/\epsilon,-1/\epsilon^2\right)_{\epsilon\rightarrow 0} = \ssf(\iu K') = \ssf_\dd(\iu K'_\dd).
\end{equation}
Note that for a generic point $\ssp=(p,P)\in\GAM$, $\ssp\cdot\ssl_\oh = (1/p,-P/p^2)$.
We now introduce two elements of $\delta$,\footnote{Actually we could choose the points $(\sse,\sse),(\ssl_\oh,\ssl_\oh)\in\delta$ here, which at first seems more natural, however the subsequent calculations require more effort.}
\begin{equation}
(\sst,\sse), \ (\sst\cdot\ssl_\oh,\ssl_\oh) \ \in \ \delta
\end{equation}
and to each we associate a new special lattice direction into which we extend $\xi_\dd$ and $\rho_1\ldots\rho_N$, denoting the shifts by $\oT_0$ and $\oT_\oh$.
Specifically (using the properties listed in the appendix)
\begin{equation}
\begin{array}{ll}
\oT_0 \xi_\dd = \xi_\dd, \quad &\oT_\oh \xi_\dd = \xi_\dd + \iu K'_\dd,\\
\oT_0 v = v, \quad & \oT_\oh v = 1/v,\\
\oT_0 \rho_j = -\rho_j, \quad & \oT_\oh \rho_j = -e^{\iu\pi\lambda_j^\dd/K_\dd}\rho_j,\\
\oT_0 \phi_j = -\phi_j, \quad & \oT_\oh \phi_j = -\left(\sn(\xi_\dd+2\lambda_j^\dd;k_\dd)/\sn(\xi_\dd;k_\dd)\right)\phi_j,
\end{array}
\label{sd}
\end{equation}
for $j\in\{1\ldots N\}$ (again, note that in (\ref{sd}) $\iu$ denotes the imaginary unit).
Our primary interest in these special lattice directions is that they facilitate the construction of a new operator
\begin{equation}
\oX:=v \oT_0\oT_\oh\label{Fdef}
\end{equation}
which has the two properties
\begin{equation}
\oX 1 = v, \quad [\oB_j,\oX]=0, \quad j\in\{1\ldots N\}.
\end{equation}
Using these properties in (\ref{nss}) we are able to establish a rather simple relationship between the functions $f$ and $g$ appearing in the $N$-soliton solution (\ref{Q4nss}), namely 
\begin{equation}
g=\oT_0\oT_\oh f,
\end{equation}
which can also be verified directly using (\ref{sd}) and the expressions for $f$ and $g$ (\ref{Q4f}) and (\ref{Q4g}).

\section{Discussion}
We have described a technique which enables solution of the B\"acklund iteration scheme for a class of integrable lattice equations.
To express the solution after $N$ applications of the B\"acklund transformation we require that the {\it covariant extension} of the initial solution be known.
This is a natural concept emerging from the feature of the lattice equations that lattice and B\"acklund directions are distinguished only by the difference of a parameter, and therefore the technique can be said to have a manifestly {\it discrete} origin.
The relative mildness of the covariant-extendibility requirement means that this approach usefully separates two hitherto intertwined aspects of the soliton-type solutions; the choice of seed solution on the one hand, and the solution of the B\"acklund iteration on the other.

The $N$-soliton formula we have discovered holds for all equations listed by Adler, Bobenko and Suris (ABS) in \cite{abs1} and is independent of the particular form of the polynomial defining the equation.
It can be said to have {\it Hirota form}, but the technique also introduces some new features, in particular there is a commuting family of second-order linear difference operators naturally associated with the solution, and these operators essentially factorise the Hirota-type polynomial.

An explicit $N$-soliton solution of the primary model listed in \cite{abs1}, $Q4$, has been given as a particular instance of the $N$-soliton formula.
We expect this solution can be written (and directly verified) in a Cauchy-matrix form generalising results in \cite{nah}, in fact an elliptic Cauchy-matrix approach has been developed by the authors alongside the present work \cite{na}.
Direct verification of this solution through bilinearization and a Casorati determinant form as in \cite{ohti}, which would generalise results of \cite{hz}, is also likely to be possible.

\begin{acknowledgements}
The authors are grateful for the hospitality of the Isaac Newton Institute in Cambridge where this work commenced during the programme Discrete Integrable Systems.
We would like to thank Simon Ruijsenaars for stimulating discussions.
JA was supported by the Australian Research Council Discovery Grant DP 0985615.
\end{acknowledgements}

\appendix
\section{Jacobi elliptic and theta functions}
In Section \ref{Q4SOL} the equation considered and its solution are given in terms of the Jacobi elliptic and theta functions. 
Definitions of these functions we adhere to can be found in Chapter 5 of \cite{akh}. 
For the convenience of the reader this appendix recalls some of the properties of these functions.

By convention the elliptic modulus is denoted by $k$ and the half-periods of $\sn(z)=\sn(z;k)$ by $2K$ and $\iu K'$.
Increments in the argument by the half-periods result in the following behaviour
\begin{equation}
\sn(z+2K) = -\sn(z),\quad \sn(z+\iu K') = \frac{1}{k\,\sn(z)}.
\end{equation}
The function $\sn(z)$ may be written in terms of the Jacobi theta functions $\ETA(z)=\ETA(z;k)$ and $\THE(z)=\THE(z;k)$,
\begin{equation}
\sqrt{k}\,\sn(z) = \frac{\ETA(z)}{\THE(z)}. \label{set}
\end{equation}
In turn these functions satisfy the basic relations
\begin{equation}
\begin{split}
\ETA(&z+2K) = \ETA(-z) = -\ETA(z), \\
\THE(&z+2K) = \THE(-z) = \THE(z), \\
\ETA(z&+\iu K') = \iu e^{-{\iu\pi}(2z+\iu K')/4K}\THE(z),\\
\THE(z&+\iu K') = \iu e^{-{\iu\pi}(2z+\iu K')/4K}\ETA(z),
\end{split}
\label{idlist}
\end{equation}
where $\iu$ is the imaginary unit.
We also note the important addition formula
\begin{equation}
\begin{split}
\ETA(x+y)\ETA(x-y)\ETA(z+w)\ETA(z-w)&\\+\ETA(x+z)\ETA(x-z)\ETA(w+y)\ETA(w-y)&\\+\ETA(x+w)\ETA(x-w)\ETA(y+z)\ETA(y-z)& = 0,
\end{split}
\label{tti}
\end{equation}
from which others can be derived.
One identity in particular is used in the text:
\begin{equation}
\frac{1-k^2\sn(a)\sn(b)\sn(c)\sn(c+a-b)}{1-k^2\sn(a)\sn(b)\sn(c)\sn(c-a+b)} = \frac{\THE(c+a)\THE(c-b)\THE(c-a+b)}{\THE(c-a)\THE(c+b)\THE(c+a-b)}.\label{mi}
\end{equation}
To derive (\ref{mi}) we choose
\begin{equation*}
\begin{array}{ll}
w=c-(a-b)/2, \quad &x=c+(a-b)/2, \\ y=(b-a)/2, \quad &z=(a+b)/2+iK'
\end{array}
\end{equation*}
and
\begin{equation*}
\begin{array}{ll}
w=c-(a-b)/2+iK', \quad &x=c+(a-b)/2+iK', \\ y=(b-a)/2, \quad &z=(a+b)/2
\end{array}
\end{equation*}
in (\ref{tti}), leading respectively to
\begin{multline}
\THE(c+a)\THE(c-b)\ETA(c-a+b)-\THE(c-a)\THE(c+b)\ETA(c+a-b)\\=\ETA(2c)\ETA(b-a)\THE(a)\THE(b)/\ETA(c)\label{ti1}
\end{multline}
and
\begin{multline}
\THE(c+a)\THE(c-b)\THE(c-a+b)-\THE(c-a)\THE(c+b)\THE(c+a-b)\\=\ETA(2c)\ETA(b-a)\ETA(a)\ETA(b)/\THE(c),\label{ti2}
\end{multline}
which we have simplified using (\ref{idlist}).
Elimination of $\ETA(2c)\ETA(b-a)$ between (\ref{ti1}) and (\ref{ti2}) followed by the use of $\ETA(z)=\sqrt{k}\,\sn(z)\THE(z)$ from (\ref{set}) to remove the remaining occurrences of $\ETA$ results in (\ref{mi}).

\end{document}